\documentclass{article}
\usepackage{authblk}
\usepackage{amsmath,amsfonts,amssymb}
\usepackage[T1]{fontenc}
\usepackage{geometry}
\usepackage{indentfirst}
\usepackage{hyperref}

\newcommand{\arxiv}[1]{\href{https://arxiv.org/abs/#1}{arXiv:#1}}
\newcommand{\bibx}[3]{#1, ``#2'', \arxiv{#3}}
\newcommand{\bibxp}[4]{#1, ``#2'', #3, \arxiv{#4}}
\newcommand{\bibp}[3]{#1, ``#2'', #3}

\def\PL#1{Phys.\ Lett.\ {\bf#1}}\def\CMP#1{Commun.\ Math.\ Phys.\ {\bf#1}}

\def\PR#1{Phys.\ Rev.\ {\bf#1}}\def\CQG#1{Class.\ Quantum Grav.\ {\bf#1}}
\def\NP#1{Nucl.\ Phys.\ {\bf#1}}
\def\JMP#1{J.\ Math.\ Phys.\ {\bf#1}}

\def\JoP#1{J.\ Phys.\ {\bf#1}} \def\IJMP#1{Int.\ J. Mod.\ Phys.\ {\bf #1}}
\def\MPL#1{Mod.\ Phys.\ Lett.\ {\bf #1}} 
\def\PRep#1{Phys.\ Rep.\ {\bf#1}}

\def\JHEP#1{JHEP\ {\bf#1}}
\def\RMP#1{Rev.\ Mod.\ Phys.\ {\bf#1}}

\def\EPJ#1{Eur.\ Phys.\ J.\ {\bf#1}}

\title{\bf Twist for Snyder space}

\author[a]{Daniel Meljanac\thanks{Daniel.Meljanac@irb.hr}}
\author[b]{Stjepan Meljanac\thanks{meljanac@irb.hr}}
\author[c,d]{Salvatore Mignemi\thanks{smignemi@unica.it}}
\author[b]{Danijel Pikuti\'{c}\thanks{dpikutic@irb.hr}}
\author[b]{Rina \v Strajn\thanks{Rina.Strajn@irb.hr}}
\affil[a]{Division of Materials Physics, Ru\dj er Bo\v skovi\'c Institute, Bijeni\v cka cesta 54, 10002 Zagreb, Croatia}
\affil[b]{Division of Theoretical Physics, Ru\dj er Bo\v skovi\'c Institute, Bijeni\v cka cesta 54, 10002 Zagreb, Croatia}
\affil[c]{Dipartimento di Matematica e Informatica, Universit\`{a} di Cagliari, viale Merello 92, 09123 Cagliari, Italy}
\affil[d]{INFN, Sezione di Cagliari, Cittadella Universitaria, 09042 Monserrato, Italy}

\begin{document}
\maketitle

\abstract{We construct the twist operator for the Snyder space. Our starting point is a non-associative star product related to a Hermitian realisation of the noncommutative coordinates originally introduced by Snyder. The corresponding coproduct of momenta is non-coassociative. The twist is constructed using a general definition of the star product in terms of a bi-differential operator in the Hopf algebroid approach. The result is given by a closed analytical expression. We prove that this twist reproduces the correct coproducts of the momenta and the Lorentz generators. The twisted Poincar\'{e} symmetry is described by a non-associative Hopf algebra, while the twisted Lorentz symmetry is described by the undeformed Hopf algebra. This new twist might be important in the construction of different types of field theories on Snyder space.}

\section{Introduction}

Since the beginning, the research of a quantum field theory brought with it the problem of ultraviolet divergences and it was already Heisenberg who proposed the idea of noncommutative spaces as a possible solution \cite{HtoP}. The idea was formulated mathematically for the first time in 1947 by Snyder \cite{Snyder}.

In his paper, Snyder has shown that the introduction of a minimal unit of length necessarily leads to a noncommutative algebra of spacetime coordinates, but also that the assumption of Lorentz covariance does not impose a requirement for the spacetime to be continuous. The noncommutative coordinates can be realised as Lie algebra generators and being represented as operators that encode rotations, their spectrum is discrete, but Lorentz invariance is preserved. The momentum space can be identified with the de Sitter space $\sim SO(4,1)/SO(3,1)$.

Because of the success of the renormalisation theory, the idea of noncommutative spaces was forgotten for many years, until it was rediscovered by mathematicians \cite{Connes} and appeared in the context of string theory \cite{SW99}. When noncommutative geometry resurfaced as an important field of research \cite{DFR,DFR2,DN,Szabo}, new models were introduced. These include e.g. the most simple type of a noncommutative space, known as the Moyal plane \cite{DFR2,DN,Szabo}, and the $\kappa$-Minkowski spacetime \cite{kappa,kappa2}, a Lie algebra-type deformation of ordinary phase space, which is especially interesting because of its connection to doubly special relativity \cite{gacM}.  The $\kappa$-Minkowski spacetime is also a noncommutative geometry where the momentum space is the de Sitter space, obtained from the Iwasawa decomposition of $SO(4,1)$ \cite{Majid,GLO2010} and that has deformed commutation relations between the momenta and the noncommutative coordinates, which leads to a noncommutative addition of momenta. It is a model that has been extensively studied in different directions, e.g., in defining star products \cite{FkgN,MSSG,KJMS,KJMS2,KJMS3}, the differential calculus \cite{BatistaM}, the conserved charges \cite{AgacAMT,FkgN2}, generalising the description of the $\kappa$-Poincar\'{e}-Hopf algebra \cite{GenPoin}, or classifying the differential calculi and considering the related field theories \cite{jhep15}.

Despite the renewed interest in noncommutative geometry, the emergence of new models and their study using the formalism of Hopf algebras \cite{Majid,ChariP}, less attention was given to Snyder's original proposal. Some of the work done includes the consideration of different field theories on the Snyder spacetime \cite{BSE} and its study in connection to quantum gravity phenomenology \cite{GLO2005}. Its classical and quantum aspects have been studied from a phenomenological point of view, outside of the noncommutative geometry formalism \cite{CM,CM2,mi,mi2,mi3,Toppan,LS,LS2,Gosh,Gosh2}. The model was also considered in a series of papers \cite{BM1,BM2,kappaS,kappaS2}, and the star product, coproduct and antipodes of its Hopf algebra were calculated.

Using a spacetime picture based on the concept of realisations of a noncommutative geometry, the authors of \cite{BM2} constructed the co-algebraic sector corresponding to any realisation of the Snyder model, and also defined a non-ambiguous self interacting scalar field theory for the Snyder spacetime, computing explicitly the first order correction terms of the corresponding Lagrangian. In \cite{GL,GL2}, the problem was considered from a geometrical point of view, with equivalent results to those of \cite{BM2}. Two types of scalar field theory were constructed. The first one, based on the natural momenta addition, uncovers a non-associative deformation of the Poincar\'{e} symmetries, and the second considers Snyder spacetime as a subspace of a larger noncommutative space. The restriction of the extra-dimensional scalar field theory was also discussed.

More recently, the construction of a free field theory  was considered in \cite{1701}, both for a hermitian realisation of the Snyder spacetime, showing that the free theory is equivalent to the commuatative one, and for a generalisation of the model, up to the first order in the deformation parameter. In \cite{MMTY}, a truncated form of the nonassociative and noncommutative Snyder $\phi^4$ field theory was defined and quantized using the functional method in momentum space. Different nonassociative star/cross product geometries, as well as the related field theories, have also been considered in \cite{Lust,Myl1,Myl2,KuprVass,KupSz,Szmonopoli}.

In this paper we construct the non-associative star product related to a Hermitian realisation of the noncommutative coordinates originally introduced by Snyder \cite{Snyder}. This construction is performed using the method proposed in \cite{BM2,1608,MM}. Using a general definition of the star product, defined by a bi-differential twist operator, we calculate the twist in the Hopf algebroid approach to all orders in the deformation parameter. The result is given by a closed analytical expression. This twist does not satisfy the cocycle condition, but we prove that it reproduces the star product and the coproducts of the momenta and the Lorentz generators. Also, it generates a quasi-bialgebra and quasi-bialgebroid structure related to the Snyder space. This twist might be important in the construction of different types of field theories on the Snyder space.

In section 2, the construction of the star product related to the original Snyder realisation is presented. In section 3, the corresponding twist operator is constructed. It is proved that it reproduces the coproducts of the momenta and the Lorentz generators. In section 4, the outlook and discussion are given. Two appendices, related to sections 2 and 3 respectively, which include important details of the proofs are added.

\section{From Snyder algebra to star product}

In its original formulation \cite{Snyder}, the Snyder algebra is generated by noncommutative coordinates $\hat{x}_\mu$, momenta $p_\mu$ and Lorentz generators $M_{\mu\nu}$ that satisfy the following commutation relations
\begin{equation}\begin{split}
&[\hat{x}_\mu, \hat{x}_\nu ] = i\beta M_{\mu\nu}, \quad [p_\mu, \hat{x}_\nu]= -i(\eta_{\mu\nu} +\beta p_\mu p_\nu), \quad [p_\mu,p_\nu]=0, \\
&[M_{\mu\nu}, M_{\rho\sigma} ]= i( \eta_{\mu\rho} M_{\nu\sigma} -\eta_{\mu\sigma} M_{\nu\rho} +\eta_{\nu\rho} M_{\mu \sigma} -\eta_{\nu\sigma} M_{\mu\rho}), \\
&[M_{\mu\nu}, \hat{x}_\lambda]= i(\eta_{\mu\lambda}\hat{x}_\nu -\eta_{\nu\lambda} \hat{x}_\mu), \quad [M_{\mu\nu}, p_\lambda]= i(\eta_{\mu\lambda}p_\nu -\eta_{\nu\lambda} p_\mu),
\end{split}\end{equation}
where $\beta$ is a constant of the order $1/M_{Pl}^2$, $M_{Pl}$ being the Planck mass and the metric tensor $\eta_{\mu\nu}= {\rm diag}(-1,1,...,1)$ has signature $(1,n-1)$. The Lorentz generators satisfy the standard relations and their commutation relations with the noncommutative coordinates and the momenta are also undeformed, i.e. vector like, hence Lorentz invariance is preserved, but the commutation relations of the momenta and the noncommutative coordinates are deformed.

It is possible to find a realisation of the Snyder algebra in terms of $x_\mu$ and $p_\mu$, the generators of the undeformed Heisenberg algebra \cite{BM1,BM2,1608}, which satisfy
\begin{equation}
[x_\mu,x_\nu]=0, \quad [p_\mu,p_\nu]=0, \quad [p_\mu,x_\nu]=-i\eta_{\mu\nu}.
\end{equation}
Moreover, we concentrate on a hermitian realisation given by
\begin{equation}\begin{split} \label{HermRealiz}
\hat{x}_\mu &=x_\mu +\frac{\beta}{2}(Dp_\mu + p_\mu D)=x_\mu +\beta \left(Dp_\mu -\frac{i}{2}p_\mu \right), \\
M_{\mu\nu} &=x_\mu p_\nu -x_\nu p_\mu , \quad p_\mu =-i \frac{\partial}{\partial x             ^\mu},
\end{split}\end{equation}
where we define the dilatation operator $D$ as
\begin{equation}
D=\frac{1}{2} (x\cdot p +p\cdot x)= x\cdot p-\frac{in}{2},
\end{equation}
and $x\cdot p$ denotes the scalar product, $x\cdot p=x^\alpha p_\alpha$. Hence
\begin{equation}
\hat{x}_\mu=x_\mu +\beta\left( x\cdot p p_\mu -i\frac{n+1}{2} p_\mu \right).
\end{equation}
Note that $M_{\mu \nu}^\dagger =M_{\mu \nu}$, $D^\dagger =D$ and $(\hat{x}_\mu)^\dagger =\hat{x}_\mu$.

For the undeformed Heisenberg algebra, $\mathcal{H}= \mathcal{A} \mathcal{T}$, where $\mathcal{A}$ is the enveloping algebra generated by $x_\mu$ and $\mathcal{T}$ is generated by $p_\mu$, we define the action $\triangleright$ on $\mathcal{A}$ , $\triangleright :\mathcal{H} \otimes \mathcal{A} \to \mathcal{A}$, by
\begin{equation}
x_\mu \triangleright f(x)= x_\mu f(x), \quad p_\mu \triangleright f(x) =-i\frac{\partial f(x)}{\partial x^\mu}, \quad f(x)\in \mathcal{A}.
\end{equation}

It can be shown \cite{1701,KMSS,JMP,EPJC2017,MM} that for realisations of noncommutative coordinates which can be written in the form $\hat{x}_\mu = x_\alpha \varphi^{\alpha}_{\,\;\mu} (p) +\chi_\mu (p)$, one has
\begin{equation} \label{def K i g}
e^{ik\cdot \hat{x}} \triangleright 1= e^{iK(k)\cdot x +ig(k)},
\end{equation}
and
\begin{equation} \label{def P i Q}
e^{ik\cdot \hat{x}} \triangleright e^{iq\cdot x}= e^{i\mathcal{P}(k,q)\cdot x+i \mathcal{Q}(k,q)},
\end{equation}
where eqs.\ \eqref{def K i g} and \eqref{def P i Q} are the defining relations for the functions $K(k)$, $g(k)$, $\mathcal{P}(k,q)$ and $\mathcal{Q} (k,q)$. It is easily seen that
\begin{equation}
\mathcal{P}_\mu (k,0)=K_\mu(k), \quad \mathcal{P}_\mu(0,q)=q_\mu.
\end{equation}

Writing the inverse of eq.\eqref{def K i g},
\begin{equation} \label{Kinverz}
e^{ik\cdot x} = e^{iK^{-1} (k)\cdot \hat{x}-ig(K^{-1}(k))} \triangleright 1,
\end{equation}
it follows that for any function $f(x)\in \mathcal{A}$ that can be Fourier transformed, we can define a function $\hat{f}(\hat{x}) \in \mathcal{H}$ in the following way
\begin{equation}
f(x)= \int d^nk \tilde{f}(k) e^{ik\cdot x}, \quad \hat{f}(\hat{x})= \int d^nk\, \tilde{f} (k)\, e^{iK^{-1} (k)\cdot \hat{x}-ig(K^{-1}(k))},
\end{equation}
and $\hat{f}$ will have the property
\begin{equation}
\hat{f}\triangleright 1= f(x).
\end{equation}

We then define the star product, $\star :\mathcal{A}\otimes \mathcal{A} \to \mathcal{A}$, of two functions $f(x),\, g(x) \in \mathcal{A}$ by
\begin{equation} \label{star prod preko trokutica}
f(x) \star g(x) =\hat{f}\hat{g} \triangleright 1= \hat{f} \triangleright g(x).
\end{equation}
From \eqref{Kinverz}, it also follows that the star product of two plane waves is given by
\begin{equation}\begin{split}
e^{ik\cdot x} \star e^{iq\cdot x} &= e^{iK^{-1}(k)\cdot \hat{x} -ig(K^{-1}(k))} \triangleright e^{iq\cdot x} \\
&= e^{i\mathcal{P}(K^{-1} (k),q)\cdot x +i\mathcal{Q}(K^{-1}(k),q) -ig(K^{-1}(k))}.
\end{split}\end{equation}
Noting that $g(k)=\mathcal{Q}(k,0)$, and defining the functions $\mathcal{D}_\mu (k,q)$ and $\mathcal{G}(k,q)$ by
\begin{align}
\mathcal{D}_\mu (k,q) &= \mathcal{P}_\mu(K^{-1}(k),q), \\
\mathcal{G}(k,q) &= \mathcal{Q}(K^{-1}(k),q)-\mathcal{Q}(K^{-1}(k),0),
\end{align}
one finds that this star product can be written as
\begin{equation}
e^{ik\cdot x}\star e^{iq\cdot x}=e^{i\mathcal{D}(k,q)\cdot x+i\mathcal{G}(k,q)}.
\end{equation}

The star product of two functions $f(x),\, g(x) \in \mathcal{A}$ that can be Fourier transformed is then given by
\begin{equation} \label{starprod preko DiG}
f(x) \star g(x) =\int d^nkd^nq \tilde{f} \tilde{g} e^{i\mathcal{D}(k,q)\cdot x+i\mathcal{G}(k,q)}.
\end{equation}
For the Hermitain realisation given by \eqref{HermRealiz}, it can be shown (for more details, see the appendix) that $\mathcal{D}_\mu(k,q)$ and $\mathcal{G}(k,q)$ are given by
\begin{align}
\mathcal{D}_\mu(k,q)&= \frac{1}{1-\beta k\cdot q} \left[ \left( 1-\frac{\beta k\cdot q}{1+ \sqrt{1+\beta k^2}} \right) k_\mu +\sqrt{1+\beta k^2} q_\mu \right], \\
\mathcal{G}(k,q)&= i\frac{n+1}{2} \ln (1-\beta k\cdot q).
\end{align}

Furthermore, we define the coproduct, $\Delta : \mathcal{T}\to \mathcal{T} \otimes \mathcal{T}$ as
\begin{equation}
\Delta p_\mu =\mathcal{D}_\mu (p\otimes 1,1\otimes p),
\end{equation}
from where one obtains the known expression for the coproduct for the so-called Snyder realisation \cite{BM2}
\begin{equation} \label{coprod p}
\Delta p_{\mu} = \frac{1}{1-\beta p_{\alpha}\otimes p^{\alpha}} \left(p_{\mu} \otimes 1 - \frac{\beta}
{1+\sqrt{1+\beta p^2}}\, p_{\mu}p_{\alpha} \otimes p^{\alpha} + \sqrt{1+\beta p^2} \otimes p_{\mu} \right).
\end{equation}
The addition of momenta which corresponds to the coproduct \eqref{coprod p} can be found in \cite{GL,GL2}. The antipode $S(p)$ is undeformed $S(p_\mu)= -p_\mu$. The star product for the Snyder space \eqref{starprod preko DiG} is non-associative, while the corresponding coproduct \eqref{coprod p} is non-coassociative.

\section{Twist from star product}

The star product $f\star g$ can also be written using the twist operator $\mathcal{F}$
\begin{equation} \label{starprod preko F-1}
f(x)\star g(x) = m\left( \mathcal{F}^{-1} (\triangleright \otimes \triangleright )(f(x) \otimes g(x))\right), \quad f,g\in \mathcal{A},
\end{equation}
where $m: \mathcal{A} \otimes \mathcal{H} \to \mathcal{H}$ is the multiplication map. From \eqref{starprod preko DiG}, within the Hopf algebroid setting, we can extract the form for $\mathcal{F}^{-1}$, the inverse of the twist operator \cite{JMP,EPJC2017,JMS,GGH,JPA2017,PLA2013,SIGMA2014}
\begin{equation}\begin{split}\label{F-1 preko::}
\mathcal{F}^{-1} &= \,\,: \exp\left[ i(1\otimes x^\alpha)(\Delta -\Delta_0) p_\alpha +i\mathcal{G}(p\otimes 1,1\otimes p) \right] :\\
&= \,\,: \exp\left[ i(1\otimes x^\alpha)(\Delta -\Delta_0) p_\alpha \right] : e^{i\mathcal{G}(p\otimes 1,1\otimes p)},
\end{split}\end{equation}
where $\Delta_0 p_\mu =p_\mu \otimes 1+1 \otimes p_\mu$ and $:\,:$ denotes normal ordering in which the coordinates $x_\alpha$ stand on the left of the momenta $p_\alpha$. Twist $\mathcal{F}^{-1}$ is determined by \eqref{F-1 preko::} up to the right ideal $\mathcal{I}_0$ of $\mathcal{H}$, which has the property $m\left( \mathcal{I}_0 (\triangleright \otimes \triangleright ) (f\otimes g)\right)=0$.

Another way for obtaining the twist is using the perturbative approach introduced in \cite{JMS}. The coproduct \eqref{coprod p} is expanded with respect to the deformation parameter $\beta$ as $\Delta p_\mu = \Sigma_{k=0}^{\infty} \Delta_k p_\mu$, with $\Delta_k p_\mu \propto \beta^k$. For $\chi (\beta p^2) =0$ (where the realisation of the noncommutative coordinates is written as $\hat{x}_\mu= x_\alpha \varphi^{\;\;\alpha}_\mu(\beta p^2)+ \beta p_\mu \chi (\beta p^2)$, see appendix), the corresponding twist is $\mathcal{F}_0=e^f$, where $f=\Sigma_{k=1}^{\infty} f_k$, $f_k\propto \beta^k$, and we choose $f_k \in \mathcal{T} \otimes \mathcal{L}(x) \mathcal{T}$, where $\mathcal{L}(x)$ is linear span of $x^\mu$. After writing down the most general ansatz for the $f_k$, one finds the unknown coefficients by requiring that the twist reproduces the coproduct \eqref{coprod p} order by order, in the following way
\begin{equation} \label{delta p preko F}
\Delta p_\mu =\mathcal{F}_0 \Delta_0 p_\mu \mathcal{F}^{-1}_0.
\end{equation}
After calculating $\mathcal{F}_0$, we find that $\mathcal{F}^{-1}$, corresponding to $\chi(\beta p^2)= -i\frac{n+1}{2} $, is given by
\begin{equation}\begin{split} \label{F-1 analit}
\mathcal{F}^{-1} &= \mathcal{F}_0^{-1} \left( \frac{1}{1\otimes 1 -\beta p_\alpha \otimes p^\alpha} \right)^{\frac{n+1}{2}} \\
&= \exp \left\{ i\left( \frac{1}{2} p^2 \otimes x\cdot p +\frac{1}{2} p_{\alpha} p_{\beta} \otimes x^{\alpha} p^{\beta}+ p_{\alpha} \otimes x\cdot p\, p^{\alpha} \right) \left( \frac{\ln(1+\beta p^2)}{p^2} \otimes 1 \right) \right\} \times \\
&\phantom{=}\quad  \left( \frac{1}{1\otimes 1 -\beta p_\alpha \otimes p^\alpha} \right)^{\frac{n+1}{2}}.
\end{split}\end{equation}
This expression is identical to the one in \eqref{F-1 preko::}, which can be proved order by order.

Note that, in the Hopf algebroid approach, we have another expression related to the twist operator
\begin{equation}
\mathcal{F}^{-1}= e^{-ip_\alpha \otimes x^\alpha}\, e^{ip_\gamma^W \otimes (x^\gamma +\beta x\cdot pp^\gamma)} \, e^{i\mathcal{G}(p\otimes 1,1\otimes p)},
\end{equation}
where $p_\mu^W$ is the momentum corresponding to the Weyl ordering, which is given by $p_\mu^W =K^{-1}_\mu(p)$. The inverse $K^{-1}_\mu$ of the function $K_\mu(p)$, defined in \eqref{def K i g}, can be calculated from the equation
\begin{equation}
K(K^{-1}(p))_\mu=p_\mu.
\end{equation}
For the realisation \eqref{HermRealiz}, the function $K_\mu(p)$ is given by $K_\mu =p_\mu \frac{\tan \sqrt{\beta p^2}}{\sqrt{\beta p^2}}$ \cite{BM2}, from where it follows that $K^{-1}_\mu$ is given by
\begin{equation}
p_\mu^{W}= K^{-1}_\mu (p)= p_\mu \frac{\arctan\sqrt{\beta p^2}}{\sqrt{\beta p^2}}.
\end{equation}
Generally, the following identity holds \cite{JPA2017}
\begin{equation}
:\, e^{i(1\otimes x^\alpha) (\Delta -\Delta_0 )p_\alpha} \,:\, = e^{-ip_\alpha \otimes x^\alpha}\, e^{iK_\gamma^{-1} (p) \otimes x_\beta \varphi^{\beta\gamma} (p)}.
\end{equation}

Having the form of the twist operator \eqref{F-1 analit}, it is important to check the consistency condition \eqref{delta p preko F}. Since $e^{i\mathcal{G}(p\otimes 1, 1 \otimes p)}$ commutes with $\Delta_0 p_\mu$, it is enough to calculate $\mathcal{F}_0 \Delta_0 p_\mu \mathcal{F}_0^{-1}$
\begin{equation} \label{prvi izraz u dokazu}
\mathcal{F}_0 \Delta_0 p_\mu \mathcal{F}^{-1}_0 =p_\mu \otimes 1 +\sum_{n=0}^\infty \sum_{k=0}^\infty \beta^{n-k} \frac{(-1)^{k+n}}{k!} A_{n,k} (p^{2(n-k)} \otimes 1) \, ad_{f_1}^k (1\otimes p_\mu),
\end{equation}
where
\begin{equation}\label{def-ank}
A_{n,k}= \sum_{r_1+...+r_k=n} \frac{1}{r_1r_2...r_k}
\end{equation}
and
\begin{equation}
ad_{f_1}^k(1\otimes p_\mu)= \beta^k \left( \sum_{l=0}^k c_{k-l,l}(p_\mu (p^2)^{k-l} \otimes 1) (p_\alpha \otimes p^\alpha)^l +\sum_{l=0}^k d_{k-l,l} ((p^2)^{k-l} \otimes p_\mu ) (p_\alpha \otimes p^\alpha )^l \right).
\end{equation}
The coefficients $c_{k-l,l}$ and $d_{k-l,l}$ satisfy the recursive relations \cite{1608}
\begin{align}
c_{k-l+1,l} &=lc_{k-l,l} +c(l-1) c_{k-l+1,l-1}+\frac{1}{2} d_{k-l+1,l-1}, \\
d_{k-l+1,l} &=\left( l+\frac{1}{2} \right) d_{k-l,l} +ld_{k-l+1,l-1},
\end{align}
with $c_{0,0}=0$ and $d_{0,0}=1$. The solutions of the recursive relations are
\begin{align}
c_{k-l,l} &= \sum_{s=0}^{l-1} (-1)^s \binom{l-1}{s} (l-s)^k -d_{k-l+1,l-1}, \label{def-c} \\
d_{k-l,l} &= \sum_{s=0}^l (-1)^s \binom{l}{s} \left( l-s+\frac{1}{2}\right)^k. \label{def-d}
\end{align}
The coefficients $c_{k-l,l}$ and $d_{k-l,l}$ satisfy the following identities, see appendix
\begin{align}
\sum_{n=0}^\infty \sum_{k=0}^n \frac{(-1)^{n+k}}{k!} A_{n,k} c_{k-l,l} x^{n-l} &=\frac{1+x-\sqrt{1+x}}{x}, \label{Id za c} \\
\sum_{n=0}^\infty \sum_{k=0}^n \frac{(-1)^{n+k}}{k!} A_{n,k} d_{k-l,l} &=\binom{\frac{1}{2}}{n-l}, \quad \forall n,l. \label{Id za d}
\end{align}
Inserting the expressions for $c_{k-l,l}$ and $d_{k-l,l}$ into \eqref{prvi izraz u dokazu}, and using the identities \eqref{Id za c} and \eqref{Id za d} leads exactly to \eqref{coprod p}.

Another check of consistency is to use the twist to calculate $\Delta M_{\mu\nu}$. One gets that
\begin{equation}
\Delta M_{\mu \nu} =\mathcal{F} \Delta_0 M_{\mu\nu} \mathcal{F}^{-1} =\Delta_0 M_{\mu\nu},
\end{equation}
where $\Delta_0 M_{\mu\nu}= M_{\mu\nu}\otimes 1+1\otimes M_{\mu\nu}$. The coproduct of the Lorentz generators turns out to be the primitive one, since $[ \mathcal{F},\Delta_0 M_{\mu\nu} ] =0$. The antipode $S(M_{\mu\nu})$ is also undeformed, $S(M_{\mu\nu}) =-M_{\mu\nu}$.

Because of the non-coassociativity of the coproduct, the twist for the Snyder space does not satisfy the cocycle condition.

Finally, from the star product \eqref{starprod preko F-1}, we define
\begin{equation}
\hat{f}(x,p) = m\left( \mathcal{F}^{-1}(\triangleright \otimes 1) (f(x) \otimes 1) \right), \quad \hat{f}(x,p) \triangleright 1= f(x),
\end{equation}
and $\hat{\mathcal{A}} =\left\{ \hat{f} (x,p)\right\} \subset \mathcal{H}$. The space $\hat{\mathcal{A}}$ is not an algebra. The non-associative algebra $\mathcal{A}_\star$, where the product of functions is given by \eqref{star prod preko trokutica}, \eqref{starprod preko DiG}, is defined by $\mathcal{A}_\star =\hat{\mathcal{A}} \triangleright 1$. Particularly,
\begin{equation}
\hat{x}_\mu (x,p) =m\left( \mathcal{F}^{-1} (\triangleright \otimes 1) (x_\mu \otimes 1) \right) = x_\mu +\beta \left( x\cdot pp_\mu -i\frac{n+1}{2} p_\mu\right),
\end{equation}
which gives back the Snyder realisation from which we have started and this proves the consistency of our approach.

Using the flipped twist operator $\mathcal{F}_{21}$, we define dual noncommutative coordinates $\hat{y}_\mu (x,p)$
\begin{equation}\begin{split}
\hat{y}_\mu (x,p) &= m\left( \mathcal{F}_{21}^{-1} (\triangleright \otimes 1) (x_\mu \otimes 1) \right) \\
&= \left( x_\mu + x\cdot pp_\mu \frac{\sqrt{1+\beta p^2}-1}{p^2} \right) \sqrt{1+\beta p^2} -i \frac{n+1}{2}\beta p_\mu.
\end{split}\end{equation}
They satisfy a Snyder type algebra \cite{1608}
\begin{equation}
[\hat{y}_\mu, \hat{y}_\nu ] = iM_{\mu\nu} \left( \frac{1-\sqrt{1+\beta p^2}}{p^2} \right) \sqrt{1+\beta p^2},
\end{equation}
and $\hat{y}_\mu \triangleright f(x) =f(x) \star x_\mu$. Note that $[\hat{x}_\mu, \hat{y}_\nu ] \neq 0$, as a consequence of the non-associativity of the star product. The relation between $\hat{y}_\mu$ and $\hat{x}_\mu$ is given by
\begin{equation}
\hat{y}_\mu =\left( \hat{x}^\alpha -\lambda^\alpha(p) \right) O^{-1}_{\alpha \mu} (p),
\end{equation}
from which it follows
\begin{equation}
x_\mu \star f(x) =\left( O_{\mu \alpha} (p) \triangleright f(x) \right) \star x^\alpha +\lambda_\mu (p) \triangleright f(x).
\end{equation}

\section{Outlook and discussion}

In this paper we have constructed the non-associative star product related to a Hermitian realisation of noncommutative coordinates, originally introduced by Snyder. Using a general definition of the star product, defined by a bi-differential twist operator, we have calculated the twist in the Hopf algebroid approach up to all orders in the deformation parameter. The result is given by a closed analytical expression. This twist does not satisfy the cocycle condition because of the non-coassociativity of the corresponding coproduct. However, we have proved that it reproduces the star product and the coproducts of the momenta and the Lorentz generators, as well as the Snyder realisation of the noncommutative coordinates. Even though the corresponding coproduct is non-coassociative, it satisfies the 3-cocycle condition (see appendix C). Twisted coalgebraic structures related to the Snyder space and Poincar\'{e} symmetry are quasi-bialgebroid and quasi-bialgebra respectively. Generalizations which include the antipode are under investigation.

We point out that the associativity of star products generally reflects the associativity of compositions of quantum mechanical operators. However, applications of the deformation quantization procedure to string theory sometimes require deformation in the direction of a quasi-Poisson bracket, which does not satisfy the Jacobi identity and hence leads to non-associativity. It was found that for string endpoints attached to a Dirichlet brane in a non-constant background $B$-field, the star products describing the correlation functions are both noncommutative and non-associative. Similarly, it was found that closed strings in flat non-geometric $R$-flux backgrounds $M$ also probe non-associative geometry. Another example of non-associative deformations arising in quantum mechanics and gravity in three dimensions is given by considering the dynamics of electrons in uniform distributions of magnetic charge. All of the above lead to non-associativity becoming an interesting area of research in physics in recent years \cite{Lust,Myl1,Myl2,KuprVass,KupSz,Szmonopoli, Malek}; for example, in connection to non-geometric $R$-flux backgrounds, non-associative star products have been investigated \cite{Myl1} and perturbative non-associative field theories were constructed \cite{Myl2}; the construction of non-associative Weyl star products was undertaken \cite{KuprVass}; physical consequences of non-associativity were examined \cite{Szmonopoli}, finding that momentum space is quantized and implying a coarse-graining of momentum space with a uniform monopole background. Regarding different directions of the research on non-associativity in quantum physics, the twist we have constructed in this paper could be helpful in studying the properties of the non-associative star product corresponding to the Snyder spacetime. Especially, it could be helpful in investigating the physical consequences of this non-associativity by generalizing approaches presented in \cite{jhep15, Asch}.

An example is given by the investigation of QFT in a Snyder background. The knowledge of the star product allows one to study the quantization using standard methods of noncommutative QFT. In this way, the free scalar field theory in the Hermitian realization \eqref{HermRealiz} has been developed in \cite{1701}, where it was shown that it is equivalent to the standard commutative theory. The interacting $\phi^4$ theory has been investigated in a linear approximation in $\beta$ in \cite{MMTY} and in the full $\beta$ dependence in \cite{MMTY2}. Of course, the interacting theory is  no longer equivalent to its commutative counterpart.
In particular, the noncommutativity and nonassociativity of the star product introduce several new features. For example, nonassociativity entails that the $\phi^4$ interaction term is not uniquely defined, but three inequivalent possibilities occur. Moreover, the total 4-momentum is not conserved in some loop diagrams. This fact makes the computation of these diagrams particularly difficult. A partial evaluation has been performed in \cite{MMTY2}, where it has been shown that the Snyder star product can introduce a natural regularization on at least some of the one-loop diagrams of the two-point functions. However, in the limit of vanishing incoming momenta, a variant of the UV/IR mixing of noncommutative field theory \cite{DN,Szabo,MRS} appears. It would be interesting to further investigate this subject. Finally, we notice that, in the context of QFT, the knowledge of the exact form of the twist  obtained in the present paper can be useful for the study of the statistics of identical particles in Snyder spacetime \cite{GGH,GuptaPRD}.

\section*{Acknowledgements}

We thank J. Trampeti\'c and J. You for useful comments and discussion. 
The work by S.M., D.P. and R.\v{S}. has been supported by Croatian Science Foundation under the Project No. IP-2014-09-9582 as well as by the H2020 Twinning project No. 692194, "RBI-TWINNING".

\appendix
\numberwithin{equation}{section}
\section{Construction of the star product}

Using \eqref{def P i Q}, it can be checked that \cite{1608,KMSS,MM}
\begin{equation}
e^{-i\lambda k\cdot \hat{x}} p_\mu e^{i\lambda k\cdot \hat{x}} \triangleright e^{iq\cdot x} = \mathcal{P}_\mu (\lambda k,q) e^{iq\cdot x},
\end{equation}
with $\lambda$ a real parameter. Differentiating both sides of this equation by $\lambda$ and noting that the realisation of the noncommutative coordinates can be written in the form $\hat{x}_\mu= x_\alpha \varphi^\alpha_{\,\mu} (\beta p^2) +\beta p_\mu \chi (\beta p^2)$, it follows that the function $\mathcal{P}_\mu (\lambda k,q)$ satisfies the differential equation
\begin{equation} \label{dif jed za P}
\frac{d\mathcal{P}_\mu (\lambda k,q)}{d\lambda} =k_\alpha\varphi_{\mu}^{\;\;\alpha} \left( \mathcal{P}(\lambda k,q) \right).
\end{equation}

Defining $\hat{x}^{\mu}_{(0)} =x^\alpha \varphi_{\mu}^{\;\;\alpha}$, one obtains that the following equality holds
\begin{equation} \label{jed za dif jed za Q}
 e^{-i\lambda k\cdot \hat{x}_{(0)}} p_\mu e^{i\lambda k\cdot \hat{x}} \triangleright e^{iq\cdot x} = \mathcal{P}_\mu(\lambda k,q) e^{iq\cdot x +i\mathcal{Q}(\lambda k,q)},
 \end{equation}
with $\lambda$ a real parameter. Differentiating both sides of \eqref{jed za dif jed za Q} by $\lambda$ and using \eqref{dif jed za P}, it follows that $\mathcal{Q}(\lambda k,q)$ satisfies the differential equation
\begin{equation}
\frac{d\mathcal{Q}(\lambda k,q)}{d\lambda} =k_\alpha \chi^\alpha\left( \mathcal{P}(\lambda k,q) \right),
\end{equation}
with $\mathcal{Q}(0,q)=0$ and $\chi^\alpha \equiv \beta p^\alpha \chi (\beta p^2)$.

Differential equations \eqref{dif jed za P} and \eqref{jed za dif jed za Q} can be used to calculate the functions $\mathcal{P}_\mu(k,q)$ and $\mathcal{Q} (k,q)$ for the hermitian realisation \eqref{HermRealiz}, which corresponds to $\varphi_{\mu}^{\;\;\alpha} =\delta^\alpha_{\mu} +\beta p_\mu p^\alpha$ and $\chi =-i\frac{n+1}{2}$. Having the functions $\mathcal{P}_\mu(k,q)$ and $\mathcal{Q} (k,q)$, one can calculate $\mathcal{D}_\mu (k,q)$ and $\mathcal{G}(k,q)$ to get
\begin{align}
\mathcal{P}_\mu (k,q) &= \frac{q_\mu + \left( \frac{\sin \sqrt{\beta k^2}}{\sqrt{\beta k^2}} +\frac{k\cdot q}{k^2} (\cos \sqrt{\beta k^2} -1 ) \right) k_\mu}{\cos \sqrt{\beta k^2} -\frac{k\cdot q}{k^2} \sqrt{\beta k^2} \sin \sqrt{\beta k^2}}, \\
\mathcal{Q}(k,q) &= i\frac{n+1}{2} \ln \left( \cos \sqrt{\beta k^2} -\frac{k\cdot q}{k^2} \sqrt{\beta k^2} \sin \sqrt{\beta k^2} \right).
\end{align}
From here it follows that
\begin{align}
\mathcal{D}_\mu(k,q) &= \frac{1}{1-\beta k\cdot q} \left[ \left( 1-\frac{\beta k\cdot q}{1+ \sqrt{1+\beta k^2}} \right) k_\mu +\sqrt{1+\beta k^2} q_\mu \right], \\
\mathcal{G}(k,q) &= i\frac{n+1}{2} \ln (1-\beta k\cdot q).
\end{align}

\section{Proofs of identities}

\subsection{General Identity}

In order to prove the identities \eqref{Id za d} and \eqref{Id za c}, it is helpful to first prove the general identity
\begin{equation}\label{id-gen}
\sum_{n=0}^\infty \sum_{k=0}^n \frac{(-1)^{n+k}}{k!} A_{n,k} \alpha^k x^n = (1+x)^\alpha.
\end{equation}
One starts by noticing that the term $(1+x)^\alpha$ can be written in the following way
\begin{equation}\begin{split}
(1+x)^\alpha &= e^{\alpha\ln(1+x)} = \sum_{k=0}^\infty \frac{(\alpha\ln(1+x))^k}{k!} \\
&=\sum_{k=0}^\infty \frac{\alpha^k}{k!} \left(\sum_{r=1}^\infty \frac{(-1)^{r+1} x^r}r\right)^k
=\sum_{k=0}^\infty \frac{\alpha^k}{k!} \prod_{i=1}^k \sum_{r_i=1}^\infty \frac{(-1)^{r_i+1} x^{r_i}}{r_i}
\end{split}\end{equation}
Since, for fixed $k$, the terms with $x^n$ are such that $r_1+...+r_k = n$ and $k\le n$, this can further be written as
\begin{equation}
\begin{split}
(1+x)^\alpha &= \sum_{n=0}^\infty \sum_{k=0}^n \frac{\alpha^k}{k!}
\left( \sum_{r_1+...r_k=n} \prod_{i=1}^k \frac{(-1)^{r_i+1}}{r_i} \right) x^n  \\
&= \sum_{n=0}^\infty \sum_{k=0}^n \frac{\alpha^k}{k!}
(-1)^{n+k} \left( \sum_{r_1+...r_k=n} \prod_{i=1}^k \frac1{r_i} \right) x^n  \\
&= \sum_{n=0}^\infty \sum_{k=0}^n \frac{\alpha^k}{k!}
(-1)^{n+k} A_{n,k} x^n.
\end{split}
\end{equation}
In the second equality, $\sum_{i=1}^k (r_i+1)=n+k$ is used and in the third equality, the definition of $A_{n,k}$ \eqref{def-ank}.

\subsection{Identity for $d$-s}

To prove \eqref{Id za d}, the identity satisfied by the coefficients $d_{k-l,l}$, \eqref{Id za d} is first multiplied by $x^{n-l}$ and the resulting expression is then summed over $n$ from 0 to $\infty$ to get
\begin{equation}\label{id-dx}
\sum_{n=0}^\infty \sum_{k=0}^n \frac{(-1)^{n+k}}{k!} A_{n,k} d_{k-l,l} x^{n-l} = \sqrt{1+x}.
\end{equation}
Using \eqref{def-d}, the definition of the coefficients $d$, and rearranging the left hand side of \eqref{id-dx}, one gets
\begin{equation}
\sum_{s=0}^l (-1)^s \binom ls x^{-l} \sum_{n=0}^\infty \sum_{k=0}^n \frac{(-1)^{n+k}}{k!} A_{n,k} \left(l-s+\frac12 \right)^k x^n = \sqrt{1+x},
\end{equation}
which, using the general identity \eqref{id-gen}, simplifies to
\begin{equation}
\sum_{s=0}^l (-1)^s \binom ls x^{-l}(1+x)^{l-s+\frac12} = \sqrt{1+x}.
\end{equation}
Therefore
\begin{equation}
\sum_{s=0}^l (-1)^s \binom ls x^{-l}(1+x)^{l-s} = 1.
\end{equation}
Using the binomial formula, the left hand side reduces to
\begin{equation}
x^{-l}\sum_{s=0}^l  \binom ls (-1)^s (1+x)^{l-s} = x^{-l}[-1+(1+x)]^l = 1,
\end{equation}
concluding the proof.

\subsection{Identity for $c$-s}

To prove \eqref{Id za c}, the identity satisfied by the coefficients $c$, one proceeds in a similar way as for the coefficients $d$. The left hand side of \eqref{Id za c} is given by
\begin{equation}
\sum_{n=0}^\infty \sum_{k=0}^n \frac{(-1)^{n+k}}{k!} A_{n,k}\left[
\sum_{s=0}^{l-1} (-1)^s \binom{l-1}s (l-s)^k - d_{k-l+1,l-1}
\right]x^{n-l},
\end{equation}
which, after setting $l=\tilde l + 1$, becomes
\begin{equation}
\frac1x
\sum_{n=0}^\infty \sum_{k=0}^n \frac{(-1)^{n+k}}{k!} A_{n,k}\left[
\sum_{s=0}^{\tilde l} (-1)^s \binom{\tilde l}s (\tilde l - s +1)^k - d_{k-\tilde l, \tilde l}
\right]x^{n-\tilde l}
\end{equation}
Using \eqref{id-dx}, this can further be written as
\begin{equation}
\frac1x
\sum_{s=0}^{\tilde l} (-1)^s \binom{\tilde l}s x^{-\tilde l}
\sum_{n=0}^\infty \sum_{k=0}^n \frac{(-1)^{n+k}}{k!} A_{n,k}
 (\tilde l - s +1)^k x^n
 - \frac{\sqrt{1+x}}x.
\end{equation}
The obtained expression is then simplified using the general identity \eqref{id-gen}
\begin{equation}\begin{split}
&\frac1x
\sum_{s=0}^{\tilde l} (-1)^s \binom{\tilde l}s x^{-\tilde l}
(1+x)^{\tilde l-s+1}
 - \frac{\sqrt{1+x}}x  \\ 
& \quad  = \frac{1+x}x
\sum_{s=0}^{\tilde l} (-1)^s \binom{\tilde l}s x^{-\tilde l}
(1+x)^{\tilde l-s}
 - \frac{\sqrt{1+x}}x \\
& \quad =
 \frac{1+x}x
x^{-\tilde l} \sum_{s=0}^{\tilde l}  \binom{\tilde l}s (-1)^s
(1+x)^{\tilde l-s}
 - \frac{\sqrt{1+x}}x  \\
& \quad =\frac{1+x}x
x^{-\tilde l} [-1+(1+x)]^{\tilde l}
 - \frac{\sqrt{1+x}}x \\
& \quad  = \frac{1+x - \sqrt{1+x}}x. \label{dokaz c zadnji korak}
\end{split}\end{equation}
The final expression in \eqref{dokaz c zadnji korak} is precisely the right hand side of \eqref{Id za c}, thus concluding the proof.

\section{3-cocycle condition}
A quasi-bialgebra over $\mathbb C$ is defined as an associative algebra $A$ together with the counit $\epsilon: A \to \mathbb C$,  coproduct $\Delta: A \to A \otimes A$ and an invertible element $\Phi\in A\otimes A\otimes A$, called the co-associator, with the following properties \cite{ChariP}
\begin{align}
(1\otimes\Delta)\Delta a &= \Phi^{-1} [(\Delta\otimes1)\Delta a]\Phi, \quad \forall a \in A, \\
\label{3coc}
[(\Delta\otimes1\otimes1)\Phi][(1\otimes1\otimes\Delta)\Phi] &= (\Phi\otimes1)[(1\otimes\Delta\otimes1)\Phi](1\otimes\Phi), \\
\label{qb-norm}
(\epsilon\otimes1)\Delta a &= a = (1\otimes\epsilon)\Delta a, \quad \forall a \in A, \\
\label{1e1}
(1\otimes\epsilon\otimes1)\Phi &= 1\otimes1.
\end{align}
Property \eqref{3coc} is called the 3-cocycle condition.

Starting with a bialgebra with coassociative coproduct $\Delta_0$, and counit $\epsilon$, it is possible to construct a quasi-bialgebra using twist $\mathcal F$, satisfying normalization condition, but not necessarily satisfying the cocycle condition. The coproducts are related by
\begin{equation}
\Delta a = \mathcal F \Delta_0 a \mathcal F^{-1}, \quad \forall a \in A.
\end{equation}

This construction indeed leads to a quasi-bialgebra. Property \eqref{qb-norm} trivially follows from the normalization condition and the co-associator is given by
\begin{equation}\label{coasF}
\Phi=(\mathcal F\otimes 1)[(\Delta_0\otimes1)\mathcal F][(1\otimes\Delta_0)\mathcal F^{-1}](1\otimes\mathcal F^{-1}).
\end{equation}
It trivially satisfies the property \eqref{1e1}, which also follows from the normalization condition. It is also straightforward, although tedious, to show that the remaining property of quasi-bialgebra -- the 3-cocycle condion \eqref{3coc} --  also holds. It follows directly from the construction \eqref{coasF} and from the coassociativity of the coproduct $\Delta_0$.

Similarly, one can extend this construction to quasi-bialgebroid.


\begin{thebibliography}{99}

\bibitem{HtoP} 
Letters of Heisenberg to Peierls (1930), in: Wolfgang Pauli, Scientific Correspondence, vol. II, 15, Ed. Karl von Meyenn, Springer-Verlang 1985.

\bibitem{Snyder} 
\bibp{H.S. Snyder}{Quantized Space-Time}{\PR{71}, 38 (1947)}.

\bibitem{Connes} 
\bibp{A.~Connes}{Non-commutative Differential Geometry}{Inst. des Hautes Etudes Scientifiques 62 (1986) 257}.

\bibitem{SW99} 
\bibxp{N.~Seiberg and E.~Witten}{String Theory and Noncommutative Geometry}{\JHEP{9909} 032 (1999)}{hep-th/9908142}.

\bibitem{DFR} 
\bibp{S.~Doplicher, K.~Fredenhagen and J.~E.~Roberts}{Space-time quantization induced by classical gravity}{\PL{B331}, 39 (1994)}.

\bibitem{DFR2} 
\bibxp{S.~Doplicher, K.~Fredenhagen and J.~E.~Roberts}{Space-time quantization induced by classical gravity}{\CMP{172} 187 (1995)}{hep-th/0303037}.

\bibitem{DN} 
\bibxp{M.~R.~Douglas and N.~A.~Nekrasov}{Noncommutative Field Theory}{\RMP{73} 977 (2001)}{hep-th/0106048}.

\bibitem{Szabo} 
\bibxp{R.~J.~Szabo}{Quantum Field Theory on Noncommutative Spaces}{\PRep{378} 207 (2003)}{hep-th/0109162}.

\bibitem{kappa}
\bibp{J.~Lukierski, H.~Ruegg, A.~Novicki and V.~N.~Tolstoy}{q-deformation of Poincar\'e algebra}{\PL{B264}, 331 (1991)}.

\bibitem{kappa2}
\bibp{J.~Lukierski, A.~Novicki and H.~Ruegg}{New quantum Poincar\'e algebra and $\kappa$-deformed field theory}{\PL{B293}, 344 (1992)}.

\bibitem{gacM} 
\bibxp{G.~Amelino-Camelia and S.~Majid}{Waves on Noncommutative Spacetime and Gamma-Ray Bursts}{\IJMP{A15} 4301 (2000)}{hep-th/9907110}.

\bibitem{Majid} 
\bibp{S.~Majid}{Foundation of quantum group theory}{Cambridge University Press 1995}.

\bibitem{GLO2010} 
\bibxp{F.~Girelli, E.~R.~Livine and D.~Oriti}{4d Deformed Special Relativity from Group Field Theories}{\PR{D81} 024015 (2010)}{0903.3475}.

\bibitem{FkgN} 
\bibxp{L.~Freidel, J.~Kowalski-Glikman and S.~Nowak}{From noncommutative kappa-Minkowski to Minkowski space-time}{\PL{B648} 70 (2007)}{hep-th/0612170}.

\bibitem{MSSG} 
\bibxp{S.~Meljanac, A.~Samsarov, M.~Stoji\'{c} and K.~S.~Gupta}{Kappa-Minkowski space-time and the star product realizations}{\EPJ{C53} 295 (2008)}{0705.2471}.

\bibitem{KJMS}
\bibxp{S.~Meljanac and M.~Stoji\'{c}}{New realizations of Lie algebra kappa-deformed Euclidean space}{\EPJ{C47} 531 (2006)}{hep-th/0605133}.

\bibitem{KJMS2} 
\bibxp{S.~Kre\v{s}i\'{c}-Juri\'{c}, S.~Meljanac and M.~Stoji\'{c}}{Covariant realizations of kappa-deformed space}{\EPJ{C51} 229 (2007)}{hep-th/0702215}.

\bibitem{KJMS3} 
\bibxp{S.~Meljanac and S.~Kre\v{s}i\'{c}-Juri\'{c}}{Differential structure on kappa-Minkowski space, and kappa-Poincar\'e algebra}{\IJMP{A26} 3385 (2011)}{1004.4647}.

\bibitem{BatistaM}
\bibxp{E.~Batista and S.~Majid}{Noncommutative geometry of angular momentum space $U(\mathfrak{su}(2))$}{\JMP{44} 107 (2003)}{hep-th/0205128}.

\bibitem{AgacAMT} 
\bibxp{A.~Agostini, G.~Amelino-Camelia, M.~Arzano, A.~Marciano and R.~A.~Tacchi}{Generalizing the Noether theorem for Hopf-algebra spacetime symmetries}{\MPL{A22} 1779 (2007)}{hep-th/0607221}.

\bibitem{FkgN2} 
\bibxp{L.~Freidel, J.~Kowalski-Glikman and S.~Nowak}{Field theory on $\kappa$-Minkowski space revisited: Noether charges and breaking of Lorentz symmetry}{\IJMP{A23} 2687-2718 (2008)}{0706.3658}.

\bibitem{GenPoin} 
\bibxp{D. Kova\v{c}evi\'{c}, S. Meljanac, A. Pacho\l{} and R. \v{S}trajn}{Generalized Poincare algebras, Hopf algebras and kappa-Minkowski spacetime}{\PL{B711} 122 (2012)}{1202.3305}.

\bibitem{jhep15}
\bibxp{T.~Juri\'{c}, S.~Meljanac, D.~Pikuti\'{c} and R.~\v{S}trajn}{Toward the classification of differential calculi on $\kappa$-Minkowski space and related field theories}{\JHEP{1507} 055 (2015)}{1502.02972}.

\bibitem{ChariP} V. Chari and A. Pressley, \textit{A Guide To Quantum Groups}, CUP, Cambridge (1994).

\bibitem{BSE} 
\bibxp{J.~C.~Breckenridge, T.~G.~Steele and V.~Elias}{Massless Scalar Field Theory in a Quantised Space-Time}{\CQG{12} 637-650 (1995)}{hep-th/9501108}.

\bibitem{GLO2005} 
\bibxp{F.~Girelli, E.~R.~Livine and D.~Oriti}{Deformed Special Relativity as an effective flat limit of quantum gravity}{\NP{B708} 411-433 (2005)}{gr-qc/0406100}.

\bibitem{CM}
\bibxp{L.~N.~Chang, D.~Mini\'c, N.~Okamura and T.~Takeuchi}{Exact Solution of the Harmonic Oscillator in Arbitrary Dimensions with Minimal Length Uncertainty Relations}{\PR{D65}, 125027 (2002)}{hep-th/0111181}.

\bibitem{CM2} 
\bibxp{S.~Benczik, L.~N.~Chang, D.~Mini\'c, N.~Okamura, S.~Rayyan and T.~Takeuchi}{Short Distance vs. Long Distance Physics: The Classical Limit of the Minimal Length Uncertainty Relation}{\PR{D66}, 026003 (2002)}{hep-th/0204049}.

\bibitem{mi} 
\bibxp{S.~Mignemi}{Classical and quantum mechanics of the nonrelativistic Snyder model}{\PR{D84}, 025021 (2011)}{1104.0490}.

\bibitem{mi2} 
\bibxp{S.~Mignemi and R.~\v Strajn}{Snyder dynamics in a Schwarzschild spacetime }{\PR{D90}, 044019 (2014)}{1404.6396}.


\bibitem{mi3}
\bibxp{S.~Mignemi and R.~\v Strajn}{Quantum mechanics on a curved Snyder space}{Adv. High Energy Phys. 1328284 (2016)}{1501.01447}.

\bibitem{Toppan}
\bibxp{P.~G.~Castro, R.~Kullock and F.~Toppan}{ Snyder Noncommutativity and Pseudo-Hermitian Hamiltonians from a Jordanian Twist}{\JMP{52}, 062105 (2011)}{1104.3852}

\bibitem{LS} 
\bibxp{L.~Lu and A.~Stern}{Snyder space revisited}{\NP{B854}, 894 (2011)}{1108.1832}.

\bibitem{LS2} 
\bibxp{L.~Lu and A.~Stern}{Particle Dynamics on Snyder space }{\NP{B860}, 186 (2012)}{1110.4112}.

\bibitem{Gosh} 
\bibxp{S.~Pramanik and S.~Gosh}{GUP-based and Snyder Non-Commutative Algebras, Relativistic Particle models and Deformed Symmetries: A Unified Approach}{\IJMP{A28}, 1350131 (2013)}{1301.4042}.

\bibitem{Gosh2}
\bibxp{S.~Pramanik, S.~Gosh and P.~Pal}{Conformal Invariance in noncommutative geometry and mutually interacting Snyder Particles}{\PR{D90}, 105027 (2014)}{1409.0689}.

\bibitem{BM1}
\bibxp{M.~V.~Battisti and S.~Meljanac}{Modification of Heisenberg uncertainty relations in non-commutative Snyder space-time geometry}{\PR{D79}, 067505 (2009)}{0812.3755}.

\bibitem{BM2}
\bibxp{M.~V.~Battisti and S.~Meljanac}{Scalar Field Theory on Non-commutative Snyder Space-Time}{\PR{D82}, 024028 (2010)}{1003.2108}.

\bibitem{kappaS} 
\bibxp{S.~Meljanac, D.~Meljanac, A.~Samsarov and M.~Stoji\'{c}}{Kappa-deformed Snyder spacetime}{\MPL{A25}, 579 (2010)}{0912.5087}.

\bibitem{kappaS2} 
\bibxp{S.~Meljanac, D.~Meljanac, A.~Samsarov and M.~Stoji\'{c}}{Kappa Snyder deformations of Minkowski spacetime, realizations and Hopf algebra}{\PR{D83}, 065009 (2011)}{1102.1655}.

\bibitem{GL} 
\bibxp{F.~Girelli and E.~R.~Livine}{Scalar field theory in Snyder space-time: alternatives}{\JHEP{1103}, 132 (2011)}{1004.0621}.

\bibitem{GL2} 
\bibxp{F.~Girelli and E.~R.~Livine}{Field theories with homogenous momentum space}{AIP Conf. Proc. \textbf{1196} 115 (2009)}{0910.3107}.

\bibitem{1701}
\bibxp{S.~Meljanac, D.~Meljanac, S.~Mignemi and R.~\v{S}trajn}{Quantum field theory in generalised Snyder spaces}{\PL{B768} 321 (2017)}{1701.05862}.

\bibitem{MMTY}
\bibxp{S.~Meljanac, S.~Mignemi, J.~Trampeti\'{c} and J.~You}{Nonassociative Snyder $\phi^4$ Quantum Field Theory}{\PR{D96}, 045021 (2017)}{1703.10851}.

\bibitem{Lust}
\bibxp{D.~Lust}{T-duality and closed string non-commutative (doubled) geometry}{\JHEP{1012} 084 (2010)}{1010.1361}.

\bibitem{Myl1}
\bibxp{D.~Mylonas, P.~Schupp and R.~J.~Szabo}{Membrane Sigma-Models and Quantization of Non-Geometric Flux Backgrounds}{\JHEP{1209} 012 (2012)}{1207.0926}.

\bibitem{Myl2}
\bibxp{D.~Mylonas and R.~J.~Szabo}{Nonassociative Field Theory on Non-Geometric Spaces}{Fortsch. Phys. \textbf{62} 727 (2014)}{1404.7304}.

\bibitem{KuprVass} 
\bibxp{V.~G.~Kupriyanov and D.~V.~Vassilevich}{Nonassociative Weyl star products}{\JHEP{1509} 103 (2015)}{1506.02329}.

\bibitem{KupSz}
\bibxp{V.~G.~Kupriyanov and R.~J.~Szabo}{$G_2$-structures and quantization of non-geometric M-theory backgrounds}{\JHEP{02} 099 (2017)}{1701.02574}.

\bibitem{Szmonopoli}
\bibxp{R.~J.~Szabo}{Magnetic monopoles and nonassociative deformations of quantum theory}{to be published in Journal of Physics: Conference Series}{1709.10080}.

\bibitem{1608}
\bibxp{S.~Meljanac, D.~Meljanac, S.~Mignemi and R.~\v{S}trajn}{Snyder-type spaces, twisted Poincar\'e algebra and addition of momenta}{\IJMP{A32} 1750172 (2017)}{1608.06207}.

\bibitem{KMSS} 
\bibxp{D.~Kovacevic, S.~Meljanac, A.~Samsarov and Z.~\v Skoda}{Hermitian realizations of kappa-Minkowski spacetime}{\IJMP{A30}, 1550019 (2015)}{1307.5772}.

\bibitem{JMP}
\bibxp{T.~Juri\'c, S.~Meljanac and D.~Pikuti\'c}{Realizations of $\kappa$-Minkowski space, Drinfeld twists and related symmetry algebras}{\EPJ{C75}, 528 (2015)}{1506.04955}.

\bibitem{EPJC2017}
\bibxp{D.~Meljanac, S,~Meljanac, D.~Pikuti\'c}{Families of vector-like deformed relativistic quantum phase spaces, twists and symmetries}{\EPJ{C77}, 830 (2017)}{1709.04745}.

\bibitem{MM}
\bibxp{S.~Meljanac, D.~Meljanac, F.~Mercati and D.~Pikuti\'c}{Noncommutative Spaces and Poincar\'e Symmetry}{\PL{B766}, 181 (2017)}{1610.06716}.

\bibitem{JMS}
\bibxp{T.~Juric, S.~Meljanac and R.~\v Strajn}{Twists, realizations and Hopf algebroid structure of kappa-deformed phase space}{\IJMP{A29}, 1450022 (2014)}{1305.3088}.

\bibitem{GGH}
\bibxp{T.~R.~Govindarajan, K.~S.~Gupta, E.~Harikumar, S.~Meljanac and D.~Meljanac}{Twisted Statistics in kappa-Minkowski Spacetime}{\PR{D77}, 105010 (2008)}{0802.1576}.

\bibitem{JPA2017}
\bibxp{S.~Meljanac, D.~Meljanac, A.~Pacho\l, D.~Pikuti\'c}{Remarks on simple interpolation between Jordanian twists}{\JoP{A50}, no.26, 265201 (2017)}{1612.07984}.

\bibitem{PLA2013}
\bibxp{T.~Juri\'c, S.~Meljanac, R.~\v{S}trajn}{$\kappa$-Poincar\'e-Hopf algebra and Hopf algebroid structure of phase space from twist}{\PL{A377}, 2472-2476 (2013)}{1303.0994}.

\bibitem{SIGMA2014}
\bibxp{T.~Juri\'c, D.~Kova\v{c}evi\'c, S.~Meljanac}{$\kappa$-Deformed Phase Space, Hopf Algebroid and Twisting}{SIGMA {\bf10}, 106 (2014)}{1402.0397}.

\bibitem{Malek}
\bibxp{M.~Gunaydin, D.~Lust, E.~Malek}{Non-associativity in non-geometric string and M-theory backgrounds, the algebra of octonions, and missing momentum modes}{\JHEP{1611}, 027 (2016)}{1607.06474}.

\bibitem{Asch}
\bibxp{P.~Aschieri, A.~Borowiec and A.~Pacho\l}{Observables and Dispersion Relations in k-Minkowski Spacetime}{\JHEP{1710}, 152 (2017)}{1703.08726}.

\bibitem{MMTY2}
\bibx{S.~Meljanac, S.~Mignemi, J.~Trampeti\'{c} and J.~You}{UV/IR Mixing in Nonassociative Snyder $\phi^4$ Theory}{1711.09639}.

\bibitem{MRS}
\bibxp{S.~Minwalla, M.~Van Raamsdonk and N.~Seiberg}{Noncommutative Perturbative Dynamics}{\JHEP {0002},  020 (2000)}{hep-th/9912072}.

\bibitem{GuptaPRD}
\bibxp{D.~Meljanac, S.~Meljanac, D.~Pikuti\'c, K.~S.~Gupta}{Twisted statistics and the structure of Lie-deformed Minkowski spaces}{\PR{D96}, 105008 (2017)}{1703.09511}.
\end{thebibliography}
\end{document}